\documentclass[aps,twocolumn,longbibliography]{revtex4-1}
\usepackage{amsmath, amssymb}
\usepackage{amsbsy, latexsym, amsfonts, amsthm}
\usepackage{mathtools}
\usepackage[colorlinks,citecolor=blue,linkcolor=blue,urlcolor=blue]{hyperref}

\def\beq{\begin{equation}}
\def\be{\begin{equation}}
\def\ee{\end{equation}}
\def\ba{\begin{eqnarray}}
\def\ea{\end{eqnarray}}

\newcommand{\dummy}{\rule{0in}{0in}}

\newcommand{\Nl}{\overset{\,\scriptscriptstyle N\!}{\ell}}
\newcommand{\Nn}{\overset{\,\scriptscriptstyle N\!}{ {n}}}

\theoremstyle{plain}
\newtheorem{theorem}{Theorem}

\newcommand{\startproof}{\setlength{\parindent}{0in}\textbf{Proof.} }
\newcommand{\finishproof}{\hfill $\blacksquare$ \\}

\begin{document}

\title{The accidental flatness constraint does not mean a wrong classical limit}

\author{Jonathan Engle${}^{a}$, Carlo Rovelli${}^{b}$}
\affiliation{${}^{a}$ Department of Physics, Florida Atlantic University, 777 Glades Road, Boca Raton, FL 33431, USA\\ 
${}^{b}$ Aix Marseille University, Universit\'e de Toulon, CNRS, CPT, 13288 Marseille, France,\\ 
Perimeter Institute, 31 Caroline Street North, Waterloo, Ontario, Canada, N2L 2Y5,\\ 
The Rotman Institute of Philosophy, 1151 Richmond St.~N London, Ontario, Canada, N6A 5B7.}

\begin{abstract}
\noindent 
We shed some light on the reason why the accidental flatness constraint appears in certain limits of the  amplitudes of covariant loop quantum gravity.   We show why this constraint is harmless, by displaying how analogous accidental constraints appear in transition amplitudes of simple systems, when certain limits are considered. 
\end{abstract}

\maketitle  

\section{Introduction}

The spinfoam amplitude of covariant loop-quantum-gravity (LQG) \cite{Livine:2007vk,Engle2008,Freidel:1998pt,Kaminski:2009fm,book:Rovelli_Vidotto_CLQG}, together with its extension with cosmological constant \cite{Han:2010uq}, defines a tentative Lorentzian quantum  theory of gravity in four dimensions.  Among the open issues of this theory is a possible objection to its viability  first raised in the literature in \cite{Bonzom:2009hw}, sharpened by a number of authors 
\cite{perini2012, hk2012, Hellmann2013a, Han2014a, Engle2020a} 
and confirmed by numerical investigations \cite{Dona2020a}: in a certain `semiclassical' limit, a `flatness constraint', or `accidental curvature constraint' appears:  
 the amplitude appears to be peaked on boundary data compatible with flat geometries only, in apparent tension with the classical limit expected from a quantum theory of gravity, which of course must include curved geometries.  

Here, building on a number of recent results, in particular the analytical and numerical investigations in  
\cite{Han2021,Asante2020,Dona2020a,Gozzini:2021kbt}, as well as original ideas proposed in \cite{Han2014a, Han:2017xwo}, 
we illustrate why the tension is only apparent, and that accidental constraints appear commonly from exchanging the order of limits.  

Since the pioneering thesis of Richard Feynman, a quantum transition amplitude can be written as a sum over paths, expressed as a limit of multiple integrals, where the limit is the refinement of a discretization of the dynamics.  In quantum gravity, the classical limit can be seen as the regime where the scale of the geometry is large compared to the Planck scale. The interplay between the two limits is subtle.  The appearance of the accidental constraint shows that if one fixes the discretization, one can find sufficiently large geometrical boundary data for which the amplitude goes wrong. But this does not conflict with the fact that for each boundary data there is a discretization for which the amplitude gives the correct result to any desired accuracy, which is what is required by consistency with the classical theory.

We illustrate this point with some explicit examples, where the logic underpinning the objection is manifestly ill conceived. 

The simple examples below show that ``accidental constraints'' analogous to the one in spinfoams are ubiquitous, especially when working with overcomplete bases, but they are harmless and they do not  indicate that the classical limit is wrong.

\section{A simple example: truncated Feynman expansion} 

The sum over history formulation of quantum theory was born in the celebrated Ph.D. Thesis by Richard Feynman  \cite{Feynman2005}.  Feynman introduces the path integral starting from the transition amplitudes of a one-dimensional system with Hamiltonian $H=H_o+V$, where $H_o$ is a free Hamiltonian and $V$ a potential, breaking the time interval in $N$ steps and inserting a resolution of the identity at each step:
\begin{align}
\nonumber
\label{qmsf}
&W(x_f, x_i; t)
:= \langle x_f | 
e^{-\frac{i}{\hbar}Ht} |x_i\rangle \\
&\dummy \quad = 
 \int  dx_n 
 \prod_{n=0}^{N-1} \langle x_{n+1} |e^{-\frac{i}{\hbar}H\frac{t}{N}}
 |x_n\rangle 
\end{align}
with $x_0=x_i$, $x_N=x_f$. Here $x$ indicates the label of a basis in the Hilbert space, and $dx_n$ the measure that gives the resolution of the identity.  The equation above is of course an identity for every $N$. 
The next step is to observe that $\epsilon := t/N$ is arbitrarily small if $N$ is sufficiently large.  In this limit, we can disregard the term quadratic in $\epsilon$ in each matrix element, and if $V$ is diagonal in $x$ we can write 
\begin{align}
\langle y|e^{-\frac{i}{\hbar}(H_o+V)\epsilon}|x\rangle &= 
\langle y |e^{-\frac{i}{\hbar}H_o\epsilon} e^{-\frac{i}{\hbar}V\epsilon}|x\rangle+O(\epsilon^2)\\
& =\langle y |e^{-\frac{i}{\hbar}H_o\epsilon}|x\rangle e^{-\frac{i}{\hbar}V(x)\epsilon}+O(\epsilon^2),
\end{align}
If the transition amplitude of $H_o$ is known, say $\langle y |e^{-\frac{i}{\hbar}H_o\epsilon}|x\rangle=e^{\frac{i}{\hbar}S_o(y,x,\epsilon)}$ then we can define a truncated amplitude 
\be
W_N(x_f, x_i; t)=
 \int  dx_n\  e^{\frac{i}{\hbar}\sum_n \epsilon L(x_{n+1},x_n, \epsilon)}.
\ee
and its limit
\be
\int [Dx(t)] e^{\frac{i}{\hbar}\int_o^t dt L(x,\dot x)}
\equiv \lim_{N\to\infty} \int  dx_n\  e^{\frac{i}{\hbar}\sum_n \epsilon L(x_{n+1},x_n, \epsilon)}.
\ee
If the above expansion in $\epsilon$ is consistent with this limit (which is not a priori obvious), this quantity gives back $W(x_f,x_i;t)$. This was Feynman's thesis. 

Now notice that since each matrix element of the truncated amplitude disregards $O(\epsilon^2)$ terms, we can equally write it as 
\be
\langle y|e^{-\frac{i}{\hbar}H\epsilon}|x\rangle =\langle y|1\!\!1-\frac{i}{\hbar}H\epsilon|x\rangle+O(\epsilon^2).
\ee
Hence within  the desired approximation we have 
\be
W_N(x_f, x_i; t)=
 \int    dx_n 
 \prod_{n=0}^{N-1} \langle x_{n+1} |1\!\!1-\frac{i}{\hbar}H\epsilon |x_n\rangle .
\ee
For large $N$, this quantity converges to the correct transition amplitude.

Now, let us assume that we repeat the above steps, but instead of using a basis of orthogonal states $|x\rangle$ we use, instead, an over-complete basis of coherent states, for instance standard wave packets of average position $q$, average momentum $p$ and width (in position space) $\sigma$, which we denote $|q,p\rangle$. 

We can repeat the steps above, obtaining a truncated transition amplitude of the form 
\begin{align}\nonumber
&W_N(q_f,p_f; q_i,p_i; t)   \\
 &\quad = \int \ \frac{dq_n dp_n}{\pi}
 \prod_{n=0}^{N-1} \langle q_{n+1}, p_{n+1}| 1\!\!1-\frac{i}{\hbar}H\epsilon
 |q_{n}, p_{n}\rangle \nonumber \\
 &\quad =: \langle q_f,p_f|U_N(t)|q_i,p_i \rangle .
\end{align}
The limit of this truncation as $N\to\infty$ still gives the correct quantum transition amplitude.  But let's observe what happens if we study the classical limit of the truncated amplitude. To this aim let us study a semiclassical regime where both $\Delta q/q$ and $\Delta p/p$ are small. This can be obtained by rescaling both $q$ and $p$, in the label of the coherent states, by $\lambda$, and considering the large $\lambda$ limit. Note that, at least for the free ($V=0$) and simple harmonic ($V=\frac{1}{2}kq^2$) cases, the classical equations of motion are invariant under such a rescaling of $q$ and $p$. 
The cut-off transition amplitude for data so rescaled is 
\begin{align}
W_N(\lambda q_f,\lambda p_f; \lambda q_i, \lambda p_i; t) = \langle \lambda q_f, \lambda p_f | U_N(t) | \lambda q_i, \lambda p_i \rangle
\end{align}
If we define the annihilation operator
\begin{align}
 {a} :=  {x} + i\frac{\sigma^2}{\hbar} {p}
\end{align}
then our coherent states are eigenstates
\begin{align}
 {a} |q',p'\rangle = \left(q' + i\frac{\sigma^2}{\hbar} p'\right) |q',p'\rangle, 
\end{align}
so that
\begin{align}
&{a} |\lambda q',\lambda p'\rangle = \lambda\, \left(q'+i\frac{\ell_o^2}{\hbar} p'\right) |\lambda q', \lambda p'\rangle .
\end{align}

Now, assume that $ {H}$ is polynomial in $ {x}$ and $ {p}$.
Then since $U_N(t)$ is  polynomial in $ {H}$, $U_N(t)$ is  polynomial in $ {x}$ and $ {p}$, and so is polynomial in $ {a}$ and $ {a}^\dagger$ as well. Choosing a normal ordering, $U_N(t)$ thus takes the form
\begin{align}
U_N(t) = \sum_{\substack{\scriptscriptstyle j=0,J\\ \scriptscriptstyle k=0,K}} C_{jk} \, ( {a}^\dagger)^j  {a}^k.
\end{align}
so that 
\begin{align}
\nonumber
&W_N(\lambda q_f,\lambda p_f; \lambda q_i,\lambda p_f; t) \\
 \label{qmsfaccidental}
&= \sum_{\substack{\scriptscriptstyle j=0,J\\ \scriptscriptstyle k=0,K}} C_{jk}\,\,\langle \lambda q_f, \lambda p_f | ( {a}^\dagger)^j  {a}^k
 | \lambda q_i, \lambda p_i \rangle \\
 \nonumber
 &= \sum_{\substack{\scriptscriptstyle j=0,J\\ \scriptscriptstyle k=0,K}}C_{jk}
{\scriptscriptstyle \left(\!q_f\! -\! i\frac{\sigma^2}{\hbar}p_f\! \right)^{\dummy \! j} \left(\!q_i\! + \! i\frac{\sigma^2}{\hbar}p_i \! \right)^{\dummy \! k}}
 \lambda^{j+k}\,\,
\langle \lambda q_f, \lambda p_f | \lambda q_i, \lambda p_i \rangle \\
 \nonumber
 &  =\! e\!\!{}^{-\lambda\!{}^2\! \left[\!\!\frac{\left(q_f\!-\!q_i\right)^2}{2\sigma^2}
+\frac{\sigma^2\! \left(p_f\!-\!p_i\!\right)^2}{4\hbar^2}\!\right]}
\!\!\!\sum_{\substack{\scriptscriptstyle j=0,J\\ \scriptscriptstyle k=0,K}}\!\! C_{jk} 
{\scriptscriptstyle \left(\!q_f\! -\! i\frac{\sigma^2}{\hbar}p_f\! \right)^{\! j} \!\left(\!q_i\!+\! i\frac{\sigma^2}{\hbar}p_i\! \right)^{\! k}}   \lambda^{j+k}.
\end{align}
Because of the finiteness of the sums, this is exponentially suppressed --- i.e., $o(\lambda^{-m})$ for all positive integers $m$  --- unless
\begin{align}
\begin{split}
q_f = q_i \quad  \text{and}  \quad
p_f = p_i .
\end{split}
\end{align}
That is, the amplitude, for fixed cut-off $N$, is exponentially suppressed unless $(q_f,p_f) = (q_i,p_i)$. 

In other words: if we take the classical limit at finite $N$, we get a constraint on the boundary data that is incompatible with the classical dynamics, or any discretized version thereof. 

If instead we removes the cut-off $N$ first, we know that we get the usual quantum dynamics and so gets no such `accidental' constraint.  Importantly this does not mean that the truncated amplitude does not capture the classical dynamics.  It does, to any desired accuracy, but a given accuracy requires an appropriately large $N$. 

\section{`Accidental' constraint in the regularized transition amplitude for `half-coherent' states in case of Larmor precession}

Let us next consider a richer example which tracks closer what happens in LQG. Consider a {charged particle on a sphere, in a uniform magnetic field}. This system admits exact, normalizable states  $|\vec{L}\rangle$ (defined below) that are analogous to the half-coherent Livine-Speziale states that play a key role in the definition of the covariant LQG amplitude \cite{Livine:2007vk}.  Time evolution in such a system, both classically and quantum mechanically, exhibits the well-known Larmor precession of the angular momentum. As we shall show, if one first expands the transition amplitude as one does in spin-foams, and implements a cut-off analogous to the cut-off on the number of vertices, and then takes the classical limit prior to removing this cut-off, one obtains another example of an `accidental constraint' --- namely that angular momentum must be constant --- inconsistent with the known exact result of Larmor precession.

In a flat Euclidean 3-space with Cartesian coordinates $(x,y,z)$,
a particle with mass $m$ and electric charge $q$ is constrained to the sphere  $r\equiv\sqrt{x^2+y^2+z^2}=R$, and driven by a uniform magnetic field $\vec{B} = B \hat{z}$, where  $\hat{z}$ is the unit vector in the $z$ direction. The vector potential in the Coulomb gauge is then
\begin{align*}
\vec{A} = \frac{1}{2} \vec{B} \times \vec{r} = \frac{B}{2}\left(-y \hat{x} +x \hat{y}\right)
\end{align*}
where $\hat{x}, \hat{y}$ are the unit vectors in the $x$ and $y$ directions.
The Lagrangian is 
\begin{align*}
L &= T-U = \frac{1}{2} m \vec{v}^2 + q \vec{v}\cdot \vec{A} \\
&= \frac{1}{2} m R^2\left(\dot{\theta}^2 + \sin^2 \theta \dot{\phi}^2\right)
+\frac{1}{2}qBR^2 \sin^2 \theta \dot{\phi} 
\end{align*}
where we have used spherical coordinates $(\theta,\phi)$ for the position of the particle, related to $(x,y,z)$ in the usual way and where dot denotes time derivative. This yields the conjugate momenta
\begin{align*}
\pi_\theta\! :=\! \frac{\partial L}{\partial \dot{\theta}}
=\! mR^2 \dot{\theta} \qquad
\pi_\phi\!:= \!\frac{\partial L}{\partial \dot{\phi}}\!
= \!mR^2\! \sin^2 \!\theta\! \left( \!\dot{\phi} + \frac{qB}{2m}\right) \!.
\end{align*}
From this, one can check that 
\begin{align}
\nonumber
\vec{L} &:= \vec{r} \times \vec{\pi} :=
\vec{r} \times \left(\vec{p} + q\vec{A}\right)\\
\label{Lclassical}
&= \left(-\sin \phi \pi_\theta -\cot \theta \cos \phi \pi_\phi \right) {x}\nonumber \\
&\ \ +\left(\cos \phi \pi_\theta -\cot \theta \sin \phi \pi_\phi \right) {y}
+\pi_\phi  {z}
\end{align}
generate rotations in the usual way and so are the physically correct angular momentum components in the presence of a magnetic field.
The Hamiltonian is
\begin{align*}
H &:= \pi_\theta \dot{\theta} + \pi_\phi \dot{\phi} - L = \frac{1}{2}mR^2\left(\dot{\theta}^2 + \sin^2\theta \dot{\phi}^2\right) \\
&
= \frac{1}{2}m\vec{v}^2 = \frac{(\vec{r}\times \vec{p})^2}{2mR^2}
= \frac{\left(\vec{L}-q\vec{r}\times\vec{p}\right)^2}{2mR^2}\\
&= \frac{\vec{L}^2}{2mR^2} + \frac{qB}{2m}L_z + \frac{q^2 B^2 R^2}{8m} \sin^2 \theta .
\end{align*}
With the usual assumption that the last term is much smaller than the others, this becomes
\begin{align}
\label{Hclassical}
H = \frac{\vec{L}^2}{2mR^2} + \frac{qB}{2m}L_z 
\end{align}
which yields 
\begin{align*}
\dot{\vec{L}} = \{\vec{L}, H\} = \frac{qB}{2m}\{\vec{L},L_z\}
\end{align*}
so that, under time evolution, $\vec{L}$ rotates about the $z$-axis with the usual Larmor angular frequency $\omega := \frac{qB}{2m}$.

Quantum states can be written as wave functions on the 2-sphere $\psi(\theta,\phi)$. Equation  (\ref{Lclassical}) leads to the standard angular momentum operators $ {L}^i$ on this space. Hermiticity of these operators forces use of the usual spherical measure $\sin^2 \theta d\theta d\phi$ in defining the inner product for the Hilbert space of states  $\mathcal{H}$. Quantization of (\ref{Hclassical}) provides an unambiguous Hamiltonian operator $ {H}$. 

We define a family of normalizable coherent states in $\mathcal{H}$ which are peaked on the operators $ {L}^i$ but not peaked in $(\theta,\phi)$ --- what we call `half-coherent' or `$\vec{L}$-coherent' states.  Specifically, for each $\vec{L}' \in \mathbb{R}^3$ such that $|\vec{L}'| =: \ell \in \mathbb{N}$, 
let $|\vec{L}'\rangle$ denote the normalized simultaneous eigenstate of $\vec{ {L}}^2$ and 
$ {n} \cdot \vec{ {L}}:= (\vec{L}'/|\vec{L}'|)\cdot\vec{ {L}}$ 
with eigenvalues 
$\hbar^2 \ell(\ell+1)$ and $\hbar \ell$, respectively, with phase chosen arbitrarily.
This family of coherent states are in fact those introduced by Livine and Speziale to quantum gravity \cite{Livine:2007vk}, and give the following expectation values and uncertainties
%
%
\begin{align}
\label{exp_uncert}
&\ \ \ \ \ \ \ \ \ \ \ \langle \vec{L}' |  {L}^i |\vec{L}' \rangle = (L')^i \\
&\qquad \Delta := \langle\vec{ {L}}^2\rangle - \langle \vec{ {L}}\rangle^2 = (\Delta L_x)^2 + (\Delta L_y)^2 + (\Delta L_z)^2
= \ell 
\nonumber
\end{align}
as well as the resolution of the identity
\begin{align*}
\mathbb{I} = \sum_{\ell=0}^\infty (2\ell + 1) \int d^2  {n} |\ell  {n} \rangle \langle \ell  {n}| .
\end{align*}
Now, suppose $|\Psi(0)\rangle = |\vec{L}'\rangle =: |\ell  {n}\rangle$ at time $t=0$. Then 
$|\Psi(T)\rangle$ at time $t=T$ satisfies
\begin{align*}
 {\vec{L}}^2 |\Psi(T)\rangle
&=  {\vec{L}}^2 
e^{\frac{i T}{\hbar} \left(\frac{ {\vec{L}}^2}{2m R^2}+\omega {L_z} \right)}|\ell  {n}\rangle \\
&=  
e^{\frac{i T}{\hbar} \left(\frac{ {\vec{L}}^2}{2m R^2}+\omega {L_z} \right)}  {\vec{L}}^2 |\ell  {n}\rangle \\&
= \hbar^2 \ell(\ell+1) |\Psi(T)\rangle 
\end{align*}
and
\begin{align*}
& (R_z(T\omega) {n})\cdot  {\vec{L}} |\Psi(T)\rangle \\
&\hspace{1cm} = e^{\frac{i T}{\hbar} \frac{ {\vec{L}}^2}{2m R^2}}
(R_z(T\omega) {n})\cdot  {\vec{L}} e^{\frac{iT\omega {L_z}}{\hbar}} |\ell  {n}\rangle \\
&\hspace{1cm}= e^{\frac{i T}{\hbar} \left(\frac{ {\vec{L}}^2}{2m R^2}+\omega {L_z} \right)} 
{n}\cdot  {\vec{L}} |\ell  {n}\rangle = \hbar \ell |\Psi(T)\rangle 
\end{align*}
where $R_z(\alpha)$ denotes rotation about the $z$ axis by angle $\alpha$.
It follows that $|\Psi(T)\rangle$ equals $|R_z(T\omega) \vec{L}'\rangle$
up to a phase, so that the quantum evolution of $\vec{L}$-coherent states exactly mimics the classical evolution of $\vec{L}$.

Now let us imitate the construction leading to the spinfoam transition amplitude.   For this, let us follow the original idea in \cite{rovelli1998}, leading to a sum over two complexes. The $\vec{L}$-coherent state transition amplitude can be written
\begin{align}
\nonumber
&W(\vec{L}_f; \vec{L}_i; T)
:= \langle \vec{L}_f | 
\exp\left(\frac{i}{\hbar} T  {H}\right) |\vec{L}_i\rangle \nonumber \\
\label{larmorsf}
&\dummy \quad = \sum_{N=0}^{\infty} \frac{1}{N!}\left(\frac{iT}{\hbar}\right)^N
 \langle \vec{L}_f |  {H}^N |\vec{L}_i\rangle \\
 \nonumber
&\dummy \quad = \sum_{N=0}^{\infty} \frac{1}{N!}\left(\frac{iT}{\hbar}\right)^N
 \left(\prod_{n=1}^{N-1}\sum_{\Nl_n=0}^\infty (2\Nl_n + 1)\int d^2\Nn_n \right)
 \\&  \dummy \quad \dummy \quad \times \prod_{n=0}^{N-1} \langle \Nl_{n+1} \Nn_{n+1} |  {H}
 | \Nl_{n} \Nn_{n}\rangle 
\end{align}
where, for each $N$, $\Nl_0 \Nn_0 := L_i$ and $\Nl_N \Nn_N := L_f$. 
The above is the analogue of the spin-foam expansion in the Livine-Speziale coherent state basis \cite{Livine:2007vk}.
From comparison with \cite{rovelli1998}, the factor 
$A(\vec{L}';\vec{L}) := \langle \vec{L}' |  {H}
 |\vec{L}\rangle$
is analogous to the vertex amplitude, and $N$ analogous to the number of vertices in the spin-foam. 

Truncating the spin-foam sum to a fixed triangulation is here analogous to taking only a single term in the above sum over $N$. More precisely, to regularize the spin-foam sum, we can put a cut-off on the number of vertices $N$ to be less than some $M$. The accidental curvature constraint has been derived for fixed triangulations; because the sum over all spin-foams with number of vertices less than $M$ is finite, such a regularized sum will still yield the same constraint.  
The regularization of (\ref{larmorsf}) analogous to this is the cut-off transition amplitude
\begin{align}
\label{larmorsfcutoff}
&W_M(\vec{L}_f; \vec{L}_i; T)
:= \sum_{N=0}^{M} \frac{1}{N!}\left(\frac{iT}{\hbar}\right)^N
 \langle \vec{L}_f |  {H}^N |\vec{L}_i\rangle \\
 \nonumber
 &\dummy \hspace{0.5in}= 
 \langle \vec{L}_f |
 \left(\sum_{N=0}^{M} \frac{1}{N!}\left(\frac{iT}{\hbar}\right)^N 
  {H}^N\right) |\vec{L}_i\rangle
  \\
 \nonumber
 &\dummy \hspace{0.5in}
 =: \langle \vec{L}_f |
 U_M(T) |\vec{L}_i\rangle 
\end{align}
where $U_M(T)$ is a `cut-off time evolution operator'.

Consider now, for each $\vec{L}_o = \ell_o  {n}_o$, the one parameter family of coherent states 
$|\lambda \vec{L}_o\rangle$.  From (\ref{exp_uncert}), we have the relative uncertainty
\begin{align*}
\frac{\Delta}{\langle  {\vec{L}}^2 \rangle}
:= \frac{(\Delta  {L}_x)^2+(\Delta  {L}_y)^2+(\Delta  {L}_x)^z}{\langle  {\vec{L}}^2 \rangle} = \frac{1}{\lambda \ell_o + 1}
\end{align*}
which goes to zero as $\lambda \rightarrow \infty$.  For this reason, the $\lambda\rightarrow \infty$ limit of such states is often taken as a classical limit. 
Furthermore, the flow $(\lambda, \vec{L}_o) \mapsto \lambda \vec{L}_o$,
underlying these families of states, is a symmetry of the classical equations of motion for $\vec{L}(t)$: 
For $\vec{L}(T) = \lambda \vec{L}_f$ and $\vec{L}(0) = \lambda \vec{L}_i$,
the classical equations of motion imply
\begin{align}
\label{cl_evol}
\vec{L}_f = R_z(\omega T) \vec{L}_i
\end{align}
independent of $\lambda$.

The cut-off transition amplitude for such families of states is 
\begin{align*}
W_M(\lambda\vec{L}_f; \lambda\vec{L}_i; T)
=  \langle \lambda \vec{L}_f |
 U_M(T) | \lambda \vec{L}_i\rangle .
\end{align*}
Now, since $U_M(T)$ is polynomial in $ {H}$, which in turn is polynomial in $ {\vec{L}}$,
theorem \ref{thmone} in the appendix below implies that this expression is zero or exponentially suppressed in the classical limit 
$\lambda\rightarrow \infty$ unless
\begin{align}
\label{larmoraccid}
\vec{L}_f = \vec{L}_i.
\end{align}
This is inconsistent with the classical evolution (\ref{cl_evol}) of $\vec{L}$.
It is a spurious `accidental' constraint arising from taking the classical limit prior to removing the cut-off $M$, similar to the accidental curvature constraint arising in spin-foams due to taking the classical limit prior to removing the cut-off on the number of vertices.

However, if one removes the cut-off $M$ first,  $M \rightarrow \infty$, from the previous section, we know that one gets the usual quantum dynamics and so gets no such `accidental' constraint. 
One can see how this is possible explicitly from equation (\ref{bound}) in the proof of theorem \ref{thmone} in the appendix: In the limit $M \rightarrow \infty$, the upper limit $N$ of the sum multiplying the suppressing exponential term becomes infinite, so that an exponentially suppressing bound is no longer implied.
Again, this doesn't mean that the correct dynamics are not captured for finite $M$, but rather that, for a given desired degree of accuracy, $M$ must be sufficiently large.

The accidental constraint in this case, as well as in the previous example, and in spin-foams, is the result of a wrong exchange of limits. \\

\centerline{---------}

CR thanks Farshid Soltani for correcting a mistake in a calculation and Pietro Don\`a and Hal Haggard for their patience in long discussions on this topic.  JE was supported in part by NSF grants PHY-1806290 and PHY-2110234.  CR was supported by the QISS JFT grant \#61466.  

\subsection*{Appendix: Theorem regarding $\vec{L}$-coherent states}

Consider the decomposition of the Hilbert space of states of our toy model
into eigenspaces of $ {\vec{L}}^2$:
\begin{align*}
\mathcal{H}=\otimes_{\ell = 0}^{\infty} \mathcal{H}_\ell .
\end{align*}
In this appendix we use the natural isomorphism \cite{Barrett:2009gg}
\begin{align}
\label{iso}
\left( {\vec{L}}, \mathcal{H}_\ell\right) \overset{\text{nat.}}{\cong}
\otimes^{2\ell}_{\text{symm.}}\left(\frac{\hbar}{2} \vec{\sigma}, \mathcal{H}_{1/2}\right)
\end{align}
between representations of $\mathfrak{su}(2)$, where 
$\sigma_i$ are the Pauli matrices, $\mathcal{H}_{1/2}:= \mathbb{C}^2$ is the Hilbert space for the spin 1/2 representation,
and $\otimes^{2\ell}_{\text{symm.}}(\cdot)$ denotes the symmetric tensor product of $2\ell$ copies of the argument.
In terms of this isomorphism, and with appropriate choice of phase convention, 
the $\vec{L}$-coherent state $|\ell n\rangle$, 
with $n = (\sin \theta_L \cos \phi_L, \sin \theta_L \sin \phi_L, \cos \theta_L)$ is given by
\begin{align}
\label{symmcoh}
|\ell  {n}\rangle = \otimes^{2\ell}| {n}\rangle,
\quad \text{with} \quad | {n}\rangle := \left( 
\begin{array}{c} \cos(\theta_L/2) \\ e^{i\phi_L}\sin(\theta_L/2)\end{array}\right) .
\end{align}

\begin{theorem}
\label{thmone}
If $ {M}$ is any operator on $\mathcal{H}$ polynomial in $\vec{ {L}}$, then
$\langle \ell'  {n}'| {M}|\ell  {n}\rangle$ is zero unless $\ell'=\ell$, and 
is exponentially suppressed as $\ell'=\ell \rightarrow \infty$ if $ {n}'\neq {n}$.
\end{theorem}
{\startproof
Using the angular momentum commutation relations, $ {M}$ can be cast in the form
\begin{align*}
 {M} = \sum_{k,m,n=0}^{N} A_{k,m,n}  {L}_x^k  {L}_y^m  {L}_z^n
\end{align*}
for some set of coefficients $\{A_{k,m,n}\}\subset \mathbb{C}$.
Since $ {L}_i$ all commute with $ {\vec{L}}^2$, $ {M}|\ell  {n}\rangle$
is again an eigenstate of $ {\vec{L}}^2$ with eigenvalue $\hbar^2 \ell(\ell+1)$,
so that $\langle \ell'  {n}'| {M}|\ell  {n}\rangle = 0$ unless $\ell'=\ell$. 
For the case $\ell'=\ell$, we have
\onecolumngrid
\begin{align*}
&\langle \ell  {n}'| {M}|\ell  {n}\rangle
= {}^{\otimes_{2\ell}}\langle  {n}' | 
 \sum_{k,m,n=0}^{N} A_{k,m,n}  {L}_x^k  {L}_y^m  {L}_z^n
| {n}\rangle^{\otimes_{2\ell}} \\
&= 
 \!\!\left(\frac{\hbar}{2}\right)^{\dummy\hspace{-0.4em}k+m+n}\hspace{-1em} \sum_{k,m,n=0}^{N} \!\!\!\!\! A_{k,m,n}\,\,\,\, {}^{\otimes_{2\ell}}\!\langle  {n}' | 
 \left(\sum_{p=1}^{2\ell}{}^p\sigma_x\right)^k \left(\sum_{q=1}^{2\ell}{}^q\sigma_y\right)^m \left(\sum_{r=1}^{2\ell}{}^r\sigma_z\right)^n
| {n}\rangle^{\otimes_{2\ell}} \\
&= \!\!\left(\frac{\hbar}{2}\right)^{\dummy\hspace{-0.4em}k+m+n}\hspace{-1em} \sum_{k,m,n=0}^{N} \!\!\!\!\! A_{k,m,n} 
{}^{\otimes_{2\ell}}\!\langle  {n}' | \!\!
 \left(\prod_{s=1}^k \sum_{p_s=1}^{2\ell}\!{}^{p_s}\!\sigma_x\!\!\right)\!\!
 \left(\prod_{u=1}^m \sum_{q_u=1}^{2\ell}\!{}^{q_u}\!\sigma_y\!\!\right)
 \!\!\left(\prod_{v=1}^n \sum_{r_v=1}^{2\ell}\!{}^{r_v}\!\sigma_z\!\!\right)
\!\!| {n}\rangle^{\otimes_{2\ell}} \\
&= \!\!\left(\frac{\hbar}{2}\right)^{\dummy\hspace{-0.4em}k+m+n}\hspace{-1em} \sum_{k,m,n=0}^{N} \!\!\!\!\! A_{k,m,n} \hspace{-1.4em}
\sum_{1\le \{p_s\}, \{q_u\}, \{r_v\} \le 2\ell} \hspace{-2em}
{}^{\otimes_{2\ell}}\!\langle  {n}' | \!\!
 \left(\prod_{s=1}^k {}^{p_s}\!\sigma_x\!\right)\!\!
 \left(\prod_{u=1}^m {}^{q_u}\!\sigma_y\!\right)\!\!
 \left(\prod_{v=1}^n {}^{r_v}\!\sigma_z\!\right) \!\!
| {n}\rangle^{\otimes_{2\ell}} 
\end{align*}
\twocolumngrid
where ${}^p\sigma_i$ denotes the action of $\sigma_i$ on the $p^{\text{th}}$ copy of $\mathcal{H}_{1/2}$ in the symmetrized tensor product decomposition of $\mathcal{H}_\ell$. 
Because $(\sigma_i)^a = 1$ for $a$ even, $(\sigma_i)^a =\sigma_i$ for $a$ odd, and 
$\sigma_x \sigma_y = i \sigma_z$ and cyclic permutations, each of the terms in the above sum consists in a product of powers of the factors
$\langle  {n}'| {n}\rangle$ and 
$\langle  {n}'|\sigma_i| {n}\rangle$.
Furthermore, in any term, the maximum number of factors of the form $\langle  {n}'|\sigma_i| {n}\rangle$ for some $i$ is $k+m+n$. As the total number of factors in each term is $2\ell$, it follows that each term has a minimum of $2\ell - (k+m+n)$ factors of the form $\langle  {n}'| {n}\rangle$. From the normalization of $| {n}\rangle$, $| {n}'\rangle$, the fact that each $\sigma_i$ has spectrum $\{\pm 1\}$, and the
Cauchy-Schwarz inequality, the rest of the factors have absolute value less than or equal to $1$.  This, combined with the triangle inequality, implies
\begin{align}
\label{bound}
& \left|\langle \ell  {n}'| {M}|\ell  {n}\rangle\right|
\le \\ \nonumber & \ \ \ \  \!\left(\hbar\ell\right)^{k+m+n}\!\!
\left(\sum_{k,m,n=0}^{N} \!\!\!\!\! |A_{k,m,n}|\right)
\left|\langle  {n}'| {n}\rangle\right|^{2\ell - (k+m+n)} .
\end{align}
Thus, if $ {n}'\neq {n}$,  so that $\langle  {n}'| {n}\rangle < 1$,
we have that  
$\left|\langle \ell  {n}'| {M}|\ell  {n}\rangle\right|$
is exponentially suppressed as $\ell \rightarrow \infty$.
\finishproof}

%

\providecommand{\href}[2]{#2}\begingroup\raggedright\endgroup

\end{document}